\documentclass[11pt]{article}

\usepackage[preprint]{acl}

\usepackage{times}
\usepackage{latexsym}

\usepackage[T1]{fontenc}

\usepackage[utf8]{inputenc}

\usepackage{microtype}

\usepackage{inconsolata}

\usepackage{graphicx}
\usepackage{colortbl}
\usepackage[table]{xcolor}
\usepackage{booktabs}

\usepackage{tcolorbox}
\usepackage{listings}
\tcbuselibrary{breakable, skins, listings}

\newtcblisting[
    auto counter,
    number within=section,
]{promptbox}[2][]{  
    colback=gray!5,
    colframe=gray!50,
    fonttitle=\bfseries,
    title={Prompt~\thetcbcounter: #1},
    label={#2},  
    boxrule=0.5pt,
    left=6pt,
    right=6pt,
    breakable,
    enhanced jigsaw,
    pad at break*=2mm,
    listing only,
    listing options={
        basicstyle=\ttfamily\small,
        breaklines=true,
        breakatwhitespace=false,
        breakindent=0pt,
        columns=fullflexible,
        keepspaces=true,
        showstringspaces=false,
    },
}

\definecolor{aclblue}{RGB}{200, 220, 245}   
\definecolor{darkblue}{RGB}{30, 60, 120}    

\title{CodeContests-O: Powering LLMs via Feedback-Driven Iterative Test Case Generation}

\author{
  Jianfeng Cai\footnotemark[2] \And Jinhua Zhu\thanks{Corresponding author.}\footnotemark[2] \And Ruopei Sun\footnotemark[2] \And Kangwen Zhao\footnotemark[2] \\ 
  \AND Dongyun Xue\footnotemark[2] \And Mingxiao Feng\footnotemark[2] \And Wengang Zhou\footnotemark[2] \And Houqiang Li\footnotemark[2] \\
  \AND \textnormal{\footnotemark[2]\,\,University of Science and Technology of China} \\
  \texttt{\{xiaobaicai,teslazhu\}@mail.ustc.edu.cn}
}

\begin{document}
\maketitle

\begin{abstract}
The rise of reasoning models necessitates large-scale verifiable data, for which programming tasks serve as an ideal source. However, while competitive programming platforms provide abundant problems and solutions, high-quality test cases for verification remain scarce. Existing approaches attempt to synthesize test cases using Large Language Models~(LLMs), but rely solely on the model's intrinsic generation capabilities without external feedback, frequently resulting in insufficiently diverse cases. To address this limitation, we propose a \textbf{Feedback-Driven Iterative Framework} for comprehensive test case construction. Specifically, our method leverages the LLM to generate initial test cases, executes them against known correct and incorrect solutions, and utilizes the failed results as feedback to guide the LLM in refining the test cases toward high fidelity and discriminability. We then apply this method to the CodeContests dataset to construct an optimized high-quality derivative, \textbf{CodeContests-O}. Evaluating against the entire pool of solutions~($1.1 \times 10^7$ in total), our dataset achieves an average True Positive Rate~(TPR) of $89.37\%$ and True Negative Rate~(TNR) of $90.89\%$, significantly outperforming the CodeContests and CodeContests+ by margins of $4.32\%$ and $9.37\%$, respectively. Furthermore, fine-tuning the Qwen2.5-7B model on CodeContests-O results in a $9.52\%$ improvement on LiveCodeBench (Pass@1). Experiments demonstrate the effectiveness of our framework and the quality of CodeContests-O. To support reproducibility and facilitate future research, we release the code\footnote{https://github.com/cai-jianfeng/CodeContests-O} and dataset\footnote{https://huggingface.co/datasets/caijanfeng/CodeContests-O}.
\end{abstract}
\section{Introduction}
\label{sec:introduction}

\begin{figure}[t]
  \includegraphics[width=\columnwidth]{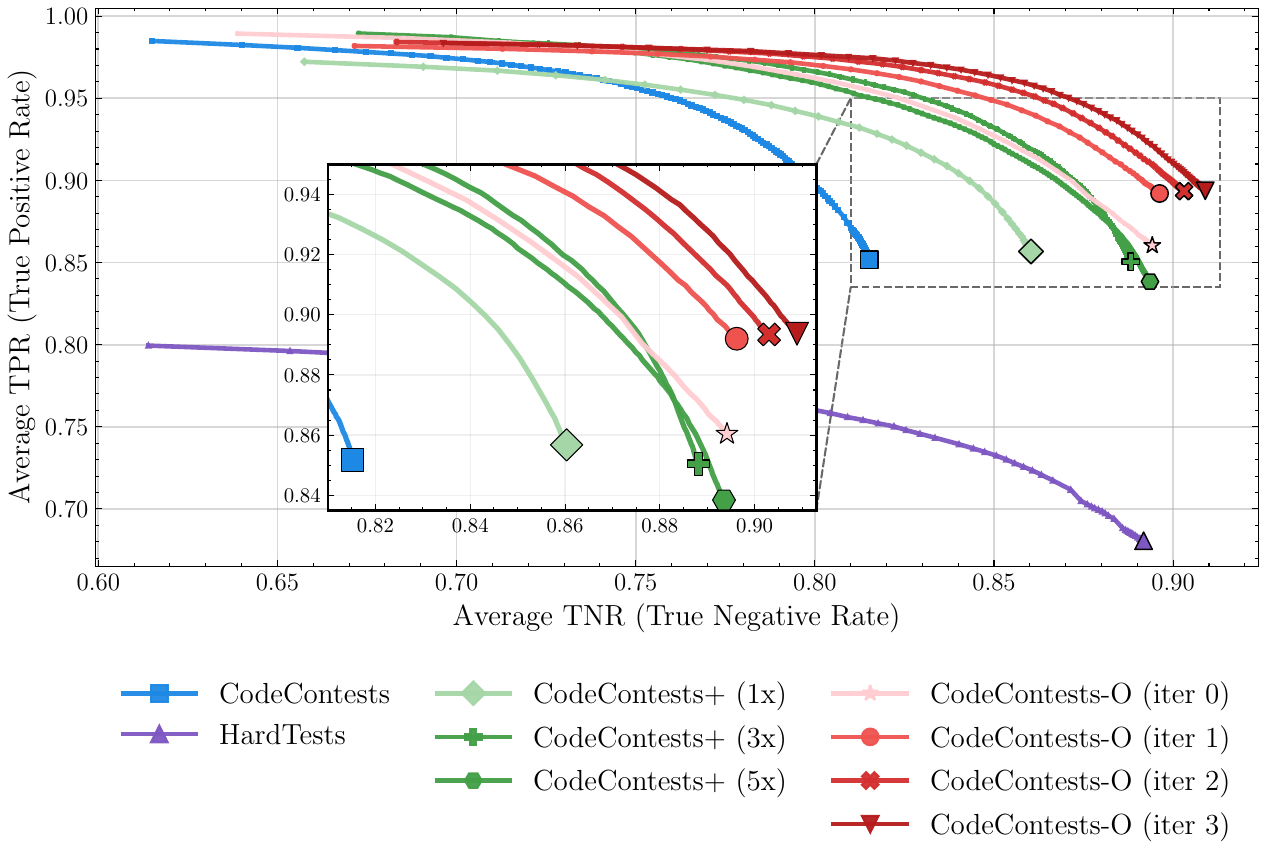}
  \caption{Pareto frontiers of TPR (True Positive Rate, proportion of correct solutions accepted, measuring test case fidelity) and TNR (True Negative Rate, proportion of incorrect solutions rejected, measuring discriminability) across different datasets. Our CodeContests-O consistently outperforms prior datasets, demonstrating superior test case quality. The iterative refinement process (iter 0-3) progressively enhances both metrics, validating the effectiveness of our feedback-driven approach.}
  \label{fig:pareto_frontier}
\end{figure}

The recent rise of reasoning-centric large language models~(LLMs)~\citep{jaech2024openai, guo2025deepseek, claude25anthropic,gemini2-525deepmind,gemini325deepmind, yang2025qwen3} marks a pivotal evolution in artificial intelligence, demonstrating unprecedented potential in solving complex logical and mathematical problems~\citep{el2025competitive, wang2025survey}. This advancement has created an urgent demand for large-scale verifiable data, for which programming tasks serve as an ideal source due to their rigorous logic and objective verifiability~\citep{lightman2023let, ni2023lever}. However, a significant bottleneck remains: while competitive programming platforms~\citep{mirzayanov2020codeforces} offer abundant problems descriptions and corresponding correct/incorrect solutions, the high-quality test cases essential for verification are often scarce or inaccessible. This deficiency makes it difficult to reliably distinguish genuine reasoning from "hallucinated" solutions that may appear correct but fail on subtle edge cases, ultimately limiting the effectiveness of model training and evaluation.

To address this scarcity, prior research has explored several avenues for automated test case generation. Early efforts relied on manual curation of small-scale benchmarks like HumanEval~\citep{chen2021evaluating} and MBPP~\citep{austin2021program}, or utilized mutation-based augmentation exemplified by AlphaCode~\citep{doi:10.1126/science.abq1158} and EvalPlus~\citep{liu2023your} to expand existing test case pools through rule-based variations. More recently, researchers have shifted toward direct LLM generation, leveraging the intrinsic knowledge of models in frameworks like CodeT~\cite{chen2022codet} and TACO~\cite{li2023taco} to synthesize test cases directly from problems. To further improve reliability, recent approaches such as CodeContests+~\citep{wang-etal-2025-codecontests} and AutoCode~\citep{zhou2025autocode} adopt a Generator-Validator paradigm, synthesizing dedicated generator and validator programs to automate the creation of large-scale test cases.

Despite these advancements, existing methods face significant limitations. They follow an open-loop paradigm that relies exclusively on LLM's intrinsic capabilities, lacking a mechanism to incorporate \textbf{objective external feedback} and \textbf{iterative refinement}. Without objective feedback signals, these methods fail to capture the diverse edge cases and dynamic failures revealed during execution against a broad solution pool. Consequently, the resulting test cases tend to be redundant or insufficiently discriminative, limiting their utility for training and evaluating robust reasoning models.

To overcome these fundamental limitations, we propose a novel framework that transforms test case synthesis from open-loop generation into a feedback-driven, iterative closed-loop process. Our approach centers on an iterative refinement loop consisting of three systematic stages. First, in \textit{Initial Test Case Generation}, we adopt a Generator-Validator paradigm where an LLM analyzes problem constraints to develop a generator program and a series of execution commands, ensuring structural integrity and diversity. Second, during \textit{Execution and Feedback Collection}, the synthesized test cases are executed against diverse solution pools. from which we capture fine-grained signals, including false positives, false negatives, and execution error logs, to quantify discriminative power. Finally, in \textit{Feedback-Guided Refinement}, the framework performs a root-cause analysis of these failures to strategically evolve the generator and its commands through a "search-and-replace" mechanism. By iteratively closing the loop between generation and execution, our method ensures that the resulting test cases are not only valid but also highly rigorous in exposing subtle algorithmic flaws.

Leveraging this framework, we then generate CodeContests-O, a high-quality verifiable code dataset derived from CodeContests~\citep{wang-etal-2025-codecontests}. By systematically applying our feedback-driven iterative refinement framework, we synthesize a large-scale test cases that achieve a high True Positive Rate (TPR) and True Negative Rate (TNR). CodeContests-O provides high-fidelity verification signals for the training and evaluation of reasoning LLMs, effectively distinguishing between genuine logical reasoning and superficial pattern matching that often fails on complex edge cases.

Empirical evaluations demonstrate the superior quality of CodeContests-O and its critical role in enhancing downstream RL training. Through extensive evaluation against the entire solution pool~($1.1 \times 10^7$ solutions in total), our iteratively refined test cases achieve an average True Positive Rate~(TPR) of $89.37\%$ and an average True Negative Rate~(TNR) of $90.89\%$, significantly outperforming original CodeContests and augmented CodeContests+. Furthermore, when employed as reward signals in RL training, CodeContests-O leads to substantial performance gains of $9.52\%$ on LiveCodeBench~\citep{jain2024livecodebench}, particularly in solving complex, logic-heavy problems. These results validate that the quality of test cases, rather than sheer quantity, is the primary driver for improving the reasoning capabilities of LLMs.

Our contributions are summarized as follows:
\begin{itemize}
    \item We propose a feedback-driven iterative generation framework that systematically synthesizes, validates, and refines test cases, effectively optimizing both fidelity and discriminability to ensure high-quality verification.
    \item We introduce CodeContests-O, a large-scale dataset with iteratively refined test cases that provides a rigorous evaluation standard to better distinguish genuine logical reasoning from superficial pattern matching.
    \item We provide a comprehensive empirical analysis demonstrating that the high-precision reward signals derived from our dataset significantly enhance downstream RL training, yielding consistent performance gains across all difficulty levels on the LiveCodeBench.
\end{itemize}
\section{Related Work}
\label{sec:related_work}

This section reviews the evolution of test case construction for reasoning LLMs, ranging from manual curation and automated synthesis to LLM-based program generation approaches.

\textbf{Manual Curation and Mutation-Based Augmentation.} Early foundational benchmarks such as HumanEval~\citep{chen2021evaluating} and MBPP~\citep{austin2021program} rely on high-quality, human-written test cases. To mitigate data contamination and keep pace with the rapid iteration of models, LiveCodeBench~\citep{jain2024livecodebench} provides a holistic, periodically updated platform by collecting problems from recent contests. While reliable, these datasets are limited in scale and struggle to capture the complexity of competitive programming. To scale test case generation beyond manual efforts, some automated approaches have integrated traditional techniques like mutation testing. Specifically, AlphaCode~\citep{doi:10.1126/science.abq1158} and EvalPlus~\citep{liu2023your} both utilize mutation-based methods to generate extensive test cases, revealing vulnerabilities in solutions that might otherwise pass simpler, manually-written tests. While these methods increase the number of test cases, they are fundamentally constrained by the initial test pool and fail to synthesize complex corner cases.

\textbf{Direct LLM Generation.} To overcome the limitations of static test cases, other research have leveraged the generative capabilities of LLMs to synthesize test cases directly. CodeT~\cite{chen2022codet} and TACO~\cite{li2023taco} leveraged the intrinsic knowledge of LLMs to directly generate test cases. Similarly, ChatTESTER~\citep{yuan2023no} and TestAug~\citep{yang2022testaug} demonstrated the potential of LLMs in augmenting test cases. CodaMosa ~\citep{lemieux2023codamosa} adopted a hybrid approach by combining LLMs with Search-Based Software Testing (SBST) to improve test coverage. However, these direct generation methods often suffer from low diversity and lack a mechanism to guarantee the validity of the generated cases.

\textbf{LLM-based Program Generation.} Recent work has explored generator-validator paradigms that leverage LLMs to synthesize dedicated programs for automated test case creation. CodeContests+~\citep{wang-etal-2025-codecontests} and AutoCode~\citep{zhou2025autocode} employ generator-validator systems to automate the test case creation process by synthesizing a dedicated generator and validator program for each problem to generate and verify test cases. Similarly, rStar-Coder~\citep{liu2025rstar} scales this paradigm by utilizing mutual verification to construct massive, verified datasets. To target complex algorithmic edge cases, HardTests~\citep{he2025hardtests} focuses on synthesizing "hacking" cases that specifically target time-limit and logic-heavy constraints. The community has also developed dedicated benchmarks like TestEval~\citep{wang2025testeval} and studies on the reliability of LLM-based test generators~\citep{cao2025can} to assess the discriminative power of these systems. However, these programmatic approaches typically rely on internal consistency or one-time checking. In contrast, our work introduces a feedback-driven iterative pipeline. By utilizing the actual execution results of known correct and incorrect solutions as a dynamic feedback signal, we guide the LLM to refine test cases iteratively to ensure high quality.
\section{Method}
\label{sec:method}

\begin{figure*}[t]
  \centering
  \includegraphics[width=0.8\linewidth]{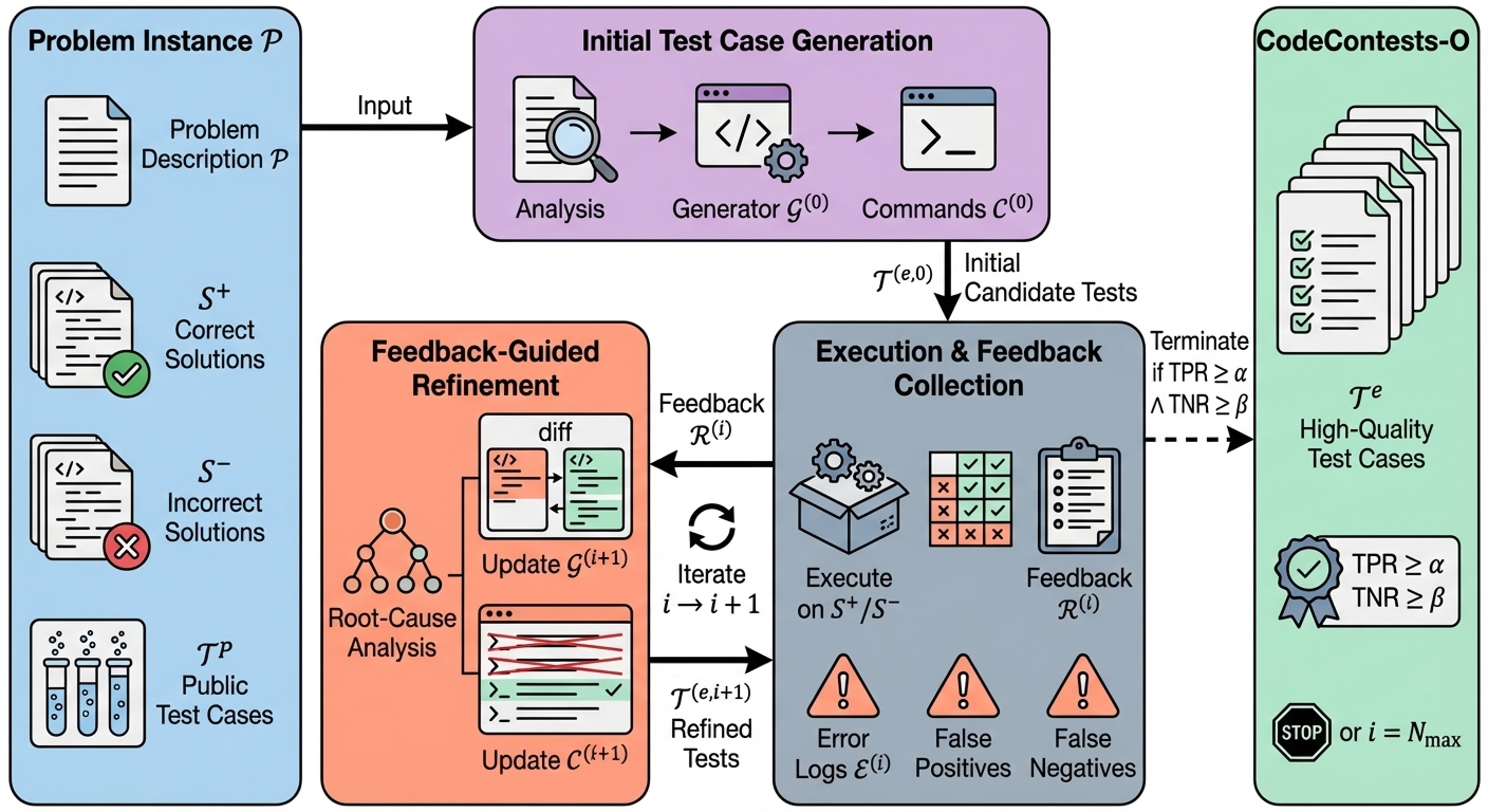}
  \caption{The overview of Feedback-Driven Iterative Test Case Generation. The framework begins with \textbf{\textit{Initial Test Case Generation}}, which analyzes the problem description $P$ to produce a generator $\mathcal{G}^{(0)}$ and commands $\mathcal{C}^{(0)}$, subsequently executing them to synthesize the initial candidate test cases $\mathcal{T}^{(e, 0)}$. This is followed by a continuous loop between \textbf{\textit{Execution and Feedback Collection}} and \textbf{\textit{Feedback-Guided Refinement}}. In each iteration $i$, test cases $\mathcal{T}^{(e, i)}$ are evaluated against $S^{+/-}$ solutions to distill a structured feedback report $\mathcal{R}^{(i)} = \{\mathcal{F}^{(i)}, \mathcal{E}^{(i)}\}$, where $\mathcal{F}^{(i)}$ identifies false positives/negatives and $\mathcal{E}^{(i)}$ captures execution error logs. The LLM then performs root-cause analysis on $\mathcal{R}^{(i)}$ to refine the generation logic. The process terminates once quality thresholds (TPR, TNR) are satisfied or the maximum iteration $N_{max}$ is reached, ultimately yielding the CodeContests-O dataset.}
  \label{fig:method_overview}
\end{figure*}

\subsection{Problem Formulation}
\label{subsec:problem_formulation}

A typical competitive programming problem is characterized by a natural language description, a reference solution, a set of public test cases, and a series of user-submitted solutions labeled by their correctness. Formally, we represent such a problem as a five-tuple $\mathcal{P} = (P, S^*, S^+, S^-, \mathcal{T}^p)$, where $P$ is the problem description, $S^*$ is the official reference solution, $S^+ = \{s_1^+, s_2^+, \ldots, s_m^+\}$ is a set of known correct solutions, $S^- = \{s_1^-, s_2^-, \ldots, s_n^-\}$ is a set of known incorrect solutions, and $\mathcal{T}^p = \{t_1^p, t_2^p, \ldots, t_k^p\}$ is a small set of public test cases provided with the problem. Each test case $t_i$ consists of an input $x_i$ and an expected output $y_i$. Our primary objective is to synthesize a high-quality, comprehensive set of test cases $\mathcal{T}^{e} = \{t_1^{e}, t_2^{e}, \ldots, t_l^{e}\}$ with superior discriminative capacity, enabling a more reliable verification process for generated code. To ensure the rigor of this verification, each test case $t_j^{e} = (x_j^{e}, y_j^{e}) \in \mathcal{T}^{e}$ must satisfy two fundamental criteria. First, it must maintain \textbf{\textit{fidelity}}: every correct solution $s_i^+ \in S^+$ must yield the expected output $y_j^{e}$ when executed with input $x_j^{e}$. Second, the generated test cases must ensure \textbf{\textit{discriminability}}: for every incorrect solution $s_i^- \in S^-$, there should exist at least one test case $t_j^{e} = (x_j^{e}, y_j^{e}) \in \mathcal{T}^{e}$ such that executing $s_i^-$ with input $x_j^{e}$ results in a discrepancy from the expected output $y_j^{e}$. By satisfying these conditions, the synthesized test cases serve as a rigorous verifier for both model training and evaluation.

\subsection{Feedback-Driven Iterative Test Case Generation}
\label{subsec:our_method}

The proposed framework for synthesizing high-quality test cases consists of three primary stages: Initial Test Case Generation, Execution and Feedback Collection, and Feedback-Guided Refinement. Figure~\ref{fig:method_overview} provides an overview of this framework.

\textbf{Initial Test Case Generation.} To establish a high-quality initial pool of test cases, we adopt a Generator-Validator paradigm consisting of the following three sequential steps.

First, we leverage the LLM to perform a comprehensive analysis of the problem $P$ by concurrently identifying structural input-output constraints and anticipating common algorithmic pitfalls, such as integer overflows, boundary conditions, or efficiency bottlenecks. This phase produces a systematic summary that guides the generation process with a thorough understanding of the task's requirements and potential edge case scenarios.

Second, based on the analytical summary, the LLM develops a dedicated generator program $\mathcal{G}^{(0)}$ to ensure the structural integrity and validity of the produced test inputs. Unlike the generation of static text, this programmatic approach utilizes external arguments to allow for precise control over data properties, enabling us to systematically adjust scales and constraints to create diverse and rigorous test inputs for various scenarios.

Finally, the LLM formulates a series of execution commands $\mathcal{C}^{(0)}$, each defining a unique set of arguments for the generator. By running the generator under these varied arguments and utilizing the reference solution $S^*$ to produce the corresponding ground-truth outputs, we construct the initial set of test cases $\mathcal{T}^{(e, 0)} = \{t_1^{(0)}, t_2^{(0)}, \dots, t_{l_0}^{(0)}\}$.

\textbf{Execution and Feedback Collection.} Following the synthesis of the initial test cases, we evaluate their quality to provide a precise signal for subsequent refinement. Specifically, we execute the candidate test cases $\mathcal{T}^{(e, i)}$ against the solution pools $S^+$ and $S^-$, recording the execution results of every correct solution $s_k^+ \in S^+$ and incorrect solution $s_m^- \in S^-$. By validating these outputs against the ground-truth outputs from the reference solution $S^*$, we quantify the discriminative power of each test case. A test case is considered more effective if it successfully validates a greater proportion of correct solutions while detecting a higher number of potential flaws in incorrect ones.

These results are then distilled into a comprehensive feedback set $\mathcal{F}^{(i)}$, which explicitly identifies false negatives (correct solutions incorrectly rejected) and false positives (incorrect solutions that bypass the test cases). These failures reveal the current test cases' limitations in distinguishing subtle algorithmic discrepancies. Furthermore, in instances where the generator $\mathcal{G}^{(i)}$ or reference solution $S^*$ fails to produce a valid output, we capture the corresponding execution error logs and stack traces, denoted as $\mathcal{E}^{(i)}$. This auxiliary information allows the LLM to diagnose underlying structural flaws or runtime crashes in the generation process. Finally, we aggregate these observations into a structured feedback report $\mathcal{R}^{(i)} = \{ \mathcal{F}^{(i)}, \mathcal{E}^{(i)} \}$, providing the LLM with the necessary context to refine and synthesize higher-quality test cases.

\begin{table*}[t]
    \centering
    \begin{tabular}{lcccc}
        \toprule
        \textbf{Dataset} & \textbf{Problems} & \textbf{Avg. Test Cases} & \textbf{Avg. $S^+$} & \textbf{Avg. $S^-$} \\
        \midrule
        CodeContests   & $13610$ & $95.81$ & $332.57$ & $649.29$ \\
        CodeContests+ (1x)  & $11690$ & $25.36$ & $373.14$ & $734.30$ \\
        CodeContests+ (3x)  & $11690$ & $61.62$ & $373.14$ & $734.30$ \\
        CodeContests+ (5x)  & $11690$ & $97.19$ & $373.14$ & $734.30$ \\
        \rowcolor{aclblue}
        CodeContests-O & $11682$ & $40.19$ & $309.23$ & $594.12$ \\
        \bottomrule
    \end{tabular}
    \caption{Comparison of statistical properties among datasets generated by different methods.}
    \label{tab:dataset_comparison}
    \vspace{-0.2cm}
\end{table*}

\textbf{Feedback-Guided Refinement.} In the final stage of each iteration $i$, we leverage the structured feedback report $\mathcal{R}^{(i)}$ to guide the LLM in a targeted refinement of the test case generation logic. This process is not a simple re-generation but a strategic evolution of the generator program $\mathcal{G}^{(i)}$ and its execution commands $\mathcal{C}^{(i)}$. 

First, the LLM performs a root-cause analysis of the failures documented in $\mathcal{R}^{(i)}$. By examining the false positives, false negatives, and execution errors $\mathcal{E}^{(i)}$, the model identifies specific weaknesses in the current generation logic. Based on these insights, the LLM updates the generator program to $\mathcal{G}^{(i+1)}$, optimizing its internal logic to capture previously overlooked boundary conditions and intricate edge cases. Simultaneously, the LLM re-designs the execution commands $\mathcal{C}^{(i)}$ into $\mathcal{C}^{(i+1)}$ to explore unexplored regions of the parameter space, ensuring that the new test cases cover complex edge cases and diverse distributions.

To implement these refinements effectively, we employ a dual-track update mechanism that operates on both the generator $\mathcal{G}^{(i)}$ and the execution commands $\mathcal{C}^{(i)}$. For the generator program, rather than rewriting the entire program, we utilize a "search-and-replace" strategy to precisely target and rectify the specific logic segments identified during the root-cause analysis. Building upon this structural update, the LLM then simultaneously updates the execution commands $\mathcal{C}^{(i)}$ by dynamically modifying the command-line argument sets. This involves selectively replacing underperforming commands to adjust the characteristics of the generated test input while adding new execution commands to probe previously unaddressed scenarios. This coordinated approach ensures that the resulting execution commands $\mathcal{C}^{(i+1)}$ are both structurally valid and increasingly rigorous.

By executing the refined command sets $\mathcal{C}^{(i+1)}$ under the updated generator $\mathcal{G}^{(i+1)}$, we produce a new batch of test inputs. These inputs are then processed by the reference solution $S^*$ to generate the corresponding ground-truth outputs, resulting in an augmented and more rigorous test case set $\mathcal{T}^{(e, i+1)} = \{t_1^{(i+1)}, t_2^{(i+1)}, \dots, t_{l_{i+1}}^{(i+1)}\}$. This new set is then fed back into the Execution and Feedback Collection phase, driving the next iteration.

\textbf{Context Compression and Checker Generation.} To prevent potential context overflow caused by lengthy iterative dialogues, we implement a context compression technique. This approach condenses information from multiple conversational turns into a single-turn summary, significantly reducing the input sequence length. Furthermore, acknowledging that programming problems often have multiple valid outputs, we employ a co-generation strategy where a checker is synthesized alongside the generator. This checker evaluates the logical consistency between a solution's output and the reference output, rather than relying on simple string matching, to ensure correctness in scenarios with non-unique valid solutions. Further details on these techniques are provided in Appendix~\ref{apx:method_details}.

This iterative refinement loop terminates when the test case set $\mathcal{T}^{e}$ reaches the target performance, where TPR $\ge \alpha$ and TNR $\ge \beta$, or fulfills the maximum iteration limit $N_{max}$. This multi-objective exit strategy ensures that the final test cases possesses robust discriminative power while maintaining computational efficiency. The complete prompt templates are provided in Appendix~\ref{sub_apx:prompt_templates}.

\subsection{Construction of CodeContests-O}
\label{subsec:dataset_construction}

To evaluate the effectiveness of the proposed framework, we apply the iterative refinement process described in Section~\ref{subsec:our_method} to construct CodeContests-O, a high-quality code dataset derived from CodeContests~\citep{wang-etal-2025-codecontests}. The construction process is organized into the following stages.

\textbf{Data Curation.} To establish a reliable foundation for the refinement process, we conduct a rigorous preprocessing of the problem set $\mathcal{D}$ (comprising individual problems $\mathcal{P} \in \mathcal{D}$) and the per-problem candidate solution pools $S^+$ and $S^-$. In particular, we first apply a set of heuristic rules as detailed in Appendix~\ref{sub_apx:problem_filter} to filter the problem set $\mathcal{D}$, ensuring that only those with appropriate algorithmic complexity and well-defined requirements are retained. Concurrently, we refine the solution pools $S^+$ and $S^-$ for each problem $\mathcal{P}$ by discarding solutions that fail to compile or execute in a standard environment. This filtering is specifically implemented by utilizing the public test cases $\mathcal{T}^p$, where we exclude any solution that fails to run successfully on all public test inputs. For the correct solution set $S^+$, we further enforce a stricter inclusion criterion requiring each solution to pass all test cases in $\mathcal{T}^p$ to guarantee its correctness.

\textbf{Dataset Synthesis.} For each curated problem $\mathcal{P}$, we execute the feedback-driven iterative test case generation framework, as detailed in Section~\ref{subsec:our_method}, using the GPT-5 model~\citep{gpt525openai} to synthesize the final CodeContests-O dataset. To ensure the reproducibility and safety of the execution process, we employ SandboxFusion\footnote{https://github.com/bytedance/SandboxFusion} as our standardized environment for running and validating all code submissions. The refinement process continues until the target thresholds are met, where we set $\alpha=0.95$ and $\beta=0.90$ for the final synthesis. These criteria ensure that the generated test cases can simultaneously validate correct solutions and effectively intercept incorrect ones. Additionally, we cap the maximum iteration limit at $N_{max}=3$ to maintain a balance between test case quality and computational cost.

\textbf{Dataset Properties.} The resulting dataset CodeContests-O comprises $11682$ unique problems. For each problem, the dataset provides a comprehensive set of test cases with an average of $40.19$ cases per problem, accompanied by a verified solution pool consisting of $309.23$ correct solutions and $594.12$ incorrect solutions. To provide a comprehensive overview of the dataset characteristics, we present a comparative analysis of statistical properties in Table~\ref{tab:dataset_comparison}, including: (1) \textbf{Problems}, which represents the count of problems; (2) \textbf{Avg. Test Cases}, reflecting the average number of tests per problem; and (3) the average size of correct and defective solution pools, \textbf{Avg. $S^+$} and \textbf{Avg. $S^-$}. It is evident that while the CodeContests and CodeContests+, rely on increasing the volume of test cases to improve the identification of defective solutions, our approach focuses on the precision and rigor of each test case. Furthermore, by excluding non-compilable or failed solutions, we ensure a higher-purity solution pool for evaluation. More detailed statistical properties of CodeContests-O are provided in Appendix~\ref{sub_apx:dataset_statistics}.
\section{Experiments}
\label{sec:experiments}

\subsection{Experimental Setup}
\label{subsec:experimental_setup}

\textbf{Compared Methods.} To evaluate the effectiveness of the proposed framework and the quality of the resulting dataset, we conduct comprehensive experiments by comparing CodeContests-O against three representative existing approaches. These include \textbf{CodeContests}~\citep{doi:10.1126/science.abq1158}, a large-scale dataset curated from competitive programming platforms; \textbf{CodeContests+}~\citep{wang-etal-2025-codecontests}, a multi-agent Generator-Validator system utilizing LLM-generated programs to synthesize test cases; and \textbf{HardTests}~\citep{he2025hardtests}, which focuses on curating adversarial instances to challenge model robustness. This systematic comparison allows us to validate our iterative refinement process against both standard data sources and specialized augmentation techniques.

\textbf{Data Quality Metrics.} The quality of synthesized test cases is quantitatively evaluated through two core metrics: \textit{True Positive Rate}~(\textbf{TPR}) and \textit{True Negative Rate}~(\textbf{TNR}). TPR reflects the \textit{fidelity} of the generated test cases, ensuring they are logically valid by measuring the proportion of known correct solutions~($S^+$) that pass the test cases without erroneous rejection. Conversely, TNR quantifies the \textit{discriminability} of the test cases by measuring the proportion of known incorrect solutions~($S^-$) that are correctly rejected. To ensure the robustness of these metrics, we utilize the entire pool of verified solutions, totaling approximately $1.1 \times 10^7$ samples for evaluation. Maximizing both metrics ensures reliable training signals and the effective removal of hallucinated solutions.

\textbf{Training and Evaluation Details.} To evaluate the practical utility of our dataset, we initialize our policy from Qwen2.5-7B~\citep{yang2024qwen2.5} and perform supervised fine-tuning~(SFT) using the CodeForces-CoTs dataset~\citep{penedo2025codeforces}. Subsequently, we apply the GRPO training paradigm~\citep{shao2024deepseekmath} on each respective dataset. We also include Qwen2.5-7B-Instruct~\citep{yang2024qwen2.5} as a reference baseline to represent the performance of a state-of-the-art model. All resulting models are evaluated on LiveCodeBench~\citep{jain2024livecodebench} using the Pass@1 metric across its three difficulty levels: Easy, Medium, and Hard. This benchmark provides a dynamic and time-sequenced stream of competitive programming problems. To ensure a fair comparison, we consistently apply identical training configurations across all compared datasets. This setup provides a direct comparison of dataset efficacy, demonstrating the superior discriminative power of CodeContests-O. Specifically, the training process is conducted on $8\times$ NVIDIA A100 GPUs using the veRL framework~\citep{sheng2025hybridflow}. For SFT, we set the maximum sequence length to $8k$, batch size to $16$, and learning rate to $1 \times 10^{-3}$. For GRPO, the maximum response length is extended to $16k$ with a batch size of $16$, a rollout size of $n=4$, and a learning rate of $1 \times 10^{-6}$.

\begin{table}[t]
    \centering
    \begin{tabular}{ccc}
        \toprule
        Dataset & TPR ($\%$) & TNR ($\%$) \\
        \midrule
        CodeContests & 85.18 & 81.52 \\
        CodeContests+ (1x) & 85.68 & 86.03 \\
        CodeContests+ (3x) & 85.05 & 88.82 \\
        CodeContests+ (5x) & 83.84 & 89.35 \\
        HardTests & 68.06 & 89.17 \\
        \arrayrulecolor{gray!100}\midrule\arrayrulecolor{black}
        \rowcolor{aclblue}
        CodeContests-O (iter 0) & 86.04 & 89.42 \\
        \rowcolor{aclblue}
        CodeContests-O (iter 1) & 89.20 & 89.62 \\
        \rowcolor{aclblue}
        CodeContests-O (iter 2) & 89.35 & 90.30 \\
        \rowcolor{aclblue}
        CodeContests-O (iter 3) & \textbf{89.37} & \textbf{90.89} \\
        \bottomrule
    \end{tabular}
    \caption{TPR and TNR results on different datasets. Here, iter 0 denotes the initial test case set produced by Initial Test Case Generation, while iter 1-3 represent the test cases synthesized in the respective $i$-th iteration of our refinement process. Our iterative refinement yields consistent metric growth, outperforming existing datasets in both fidelity and discriminative power.}
    \label{tab:tpr_tnr_result}
    \vspace{-0.2cm}
\end{table}

\begin{table*}[t]
    \centering
    \begin{tabular}{cccccc}
        \toprule
        Dataset & Pass@1 (\%) & Easy (\%) & Medium (\%) & Hard (\%) \\
        \midrule
        Qwen2.5-7B-Instruct & 29.33 & 57.52 & 23.98 & \textbf{2.70} \\
        Qwen2.5-7B (SFT) & 25.05 & 58.91 & 13.33 & 0.61 \\
        CodeContests & 27.10 & 64.66 & 13.55 & 0.77 \\
        CodeContests+ (1x) & 26.52 & 65.08 & 12.19 & 0.10 \\
        CodeContests+ (3x) & 29.17 & 67.52 & 16.42 & 0.77 \\
        CodeContests+ (5x) & 29.61 & 64.37 & 19.93 & 1.17 \\
        HardTests & 26.54 & 62.61 & 14.09 & 0.46 \\
        \rowcolor{aclblue}
        CodeContests-O & \textbf{34.57} & \textbf{70.42} & \textbf{26.56} & \textbf{2.45} \\
        \bottomrule
    \end{tabular}
    \caption{Performance comparison on LiveCodeBench across models trained with different datasets. Qwen2.5-7B~(SFT) is the model after supervised fine-tuning, used as the starting point for RL. The superior Pass@1 results achieved by our CodeContests-O demonstrate that higher reward fidelity during RL directly translates to stronger generalization in competitive programming tasks.}
    \label{tab:rl_training_result}
    \vspace{-0.2cm}
\end{table*}

\subsection{Verification of Test Case Quality}
\label{subsec:verification_of_test_case_quality}

Table~\ref{tab:tpr_tnr_result} presents the TPR and TNR results across different datasets, where CodeContests-O demonstrates superior fidelity and significantly enhanced discriminative power. Specifically, the original CodeContests exhibits a limited TNR of $81.52\%$, whereas our dataset significantly tightens the evaluation rigors by achieving a substantially higher TNR of $90.89\%$. Furthermore, while simply scaling the quantity of test cases, exemplified by CodeContests+~(3x) and~(5x), raises the TNR to $89.35\%$, it fails to maintain a high TPR, indicating that density-driven augmentation often introduces noise or redundancy. In contrast, CodeContests-O achieves a superior balance, outperforming CodeContests+~(5x) in both TPR ($89.37\%$) and TNR ($90.89\%$). Notably, while HardTests is specifically designed for adversarial robustness, it achieves a TNR of $89.17\%$ at the expense of a lower TPR ($68.06\%$), suggesting that its over-specialization on edge cases may inadvertently filter out valid solutions. This dual-metric dominance confirms that our feedback-driven iterative framework generates test cases that are not only more diverse but also more precise in identifying subtle logic errors.

To further characterize the quality distribution, we evaluate the Pareto frontiers of these datasets. As illustrated in Figure~\ref{fig:pareto_frontier}, we rank the test cases by their individual performance and aggregate them descendingly to visualize the trade-off between TPR and TNR. The Pareto frontier of CodeContests-O consistently envelopes those of the comparative datasets, demonstrating that our method provides a superior set of test cases that achieve higher fidelity at any given level of discriminative power. Moreover, the progressive expansion of the Pareto frontiers toward the top-right corner (Figure~\ref{fig:pareto_frontier}), coupled with the consistent improvement of TPR and TNR results (Table~\ref{tab:tpr_tnr_result}) from iter 0 to iter 3, validates the efficacy of our feedback-driven iterative framework. Specifically, the TPR increases from $86.04\%$ to $89.37\%$, while the TNR improves from $89.42\%$ to $90.89\%$. This dual evidence confirms that our refinement process effectively filters out low-quality cases while synthesizing more challenging ones, resulting in test cases that achieve both extensive coverage of edge cases and exceptional discriminative rigor.

\begin{figure}[t]
  \includegraphics[width=\columnwidth]{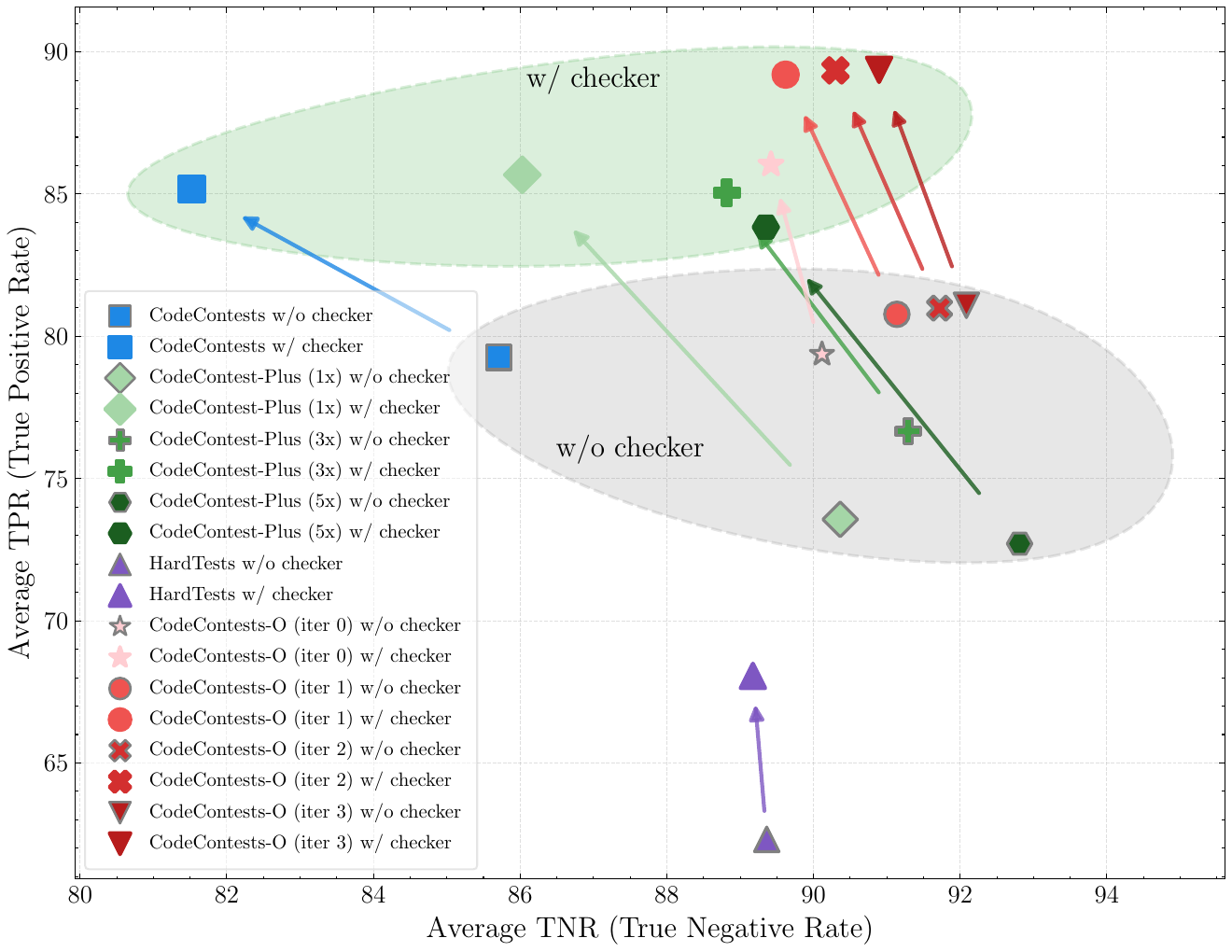}
  \caption{Effect of logic-based checker on TPR and TNR. "\textit{w/o checker}" denotes string-matching evaluation, while "\textit{w/ checker}" employs the logic-based checker. The checker consistently improves TPR across all datasets by correctly validating solutions with multiple valid outputs. CodeContests-O with checker achieves the optimal balance, demonstrating the effectiveness of our synchronized checker generation strategy.}
  \label{fig:checker}
  \vspace{-0.2cm}
\end{figure}

\subsection{RL Results with Test Cases}
\label{subsec:rl_result}

Table~\ref{tab:rl_training_result} summarizes the performance of Qwen2.5-7B models trained on different datasets. A key observation is the performance gap between the initial models: Qwen2.5-7B-Instruct achieves $29.33\%$, significantly outperforming the $25.05\%$ of our starting SFT model. This suggests that while SFT on domain-specific code data provides a foundation, it lacks the broader problem-solving proficiency of a general-purpose instruct model. However, after applying RL with our curated dataset, the model trained on CodeContests-O achieves an accuracy of $34.57\%$, representing a significant improvement of $+7.47\%$ over the original CodeContests and, crucially, surpassing the Qwen2.5-7B-Instruct.

This substantial gain underscores the practical benefits of our high-quality test cases in enhancing model reasoning capabilities. In comparison, models trained on CodeContests+ (5x) and HardTests yield accuracies of $29.61\%$ and $26.54\%$, respectively, both falling short of CodeContests-O. Furthermore, CodeContests-O consistently outperforms all compared methods across Easy, Medium, and Hard tasks. These results highlight that test case quality matters more than quantity for effective RL training, and validate our framework's ability to generate high-fidelity test cases that achieve state-of-the-art performance.

\subsection{Effect of Checker}

As shown in Figure~\ref{fig:checker}, enabling the checker consistently increases TPR while slightly reducing TNR. This is expected in multi-solution problems, where string matching without a checker incorrectly rejects semantically correct solutions, leading to false negatives. By verifying logical consistency instead of exact outputs, the checker resolves this ambiguity and improves verification fidelity. The resulting TPR-TNR trade-off yields a more balanced and reliable evaluator for downstream training.

\section{Conclusion}
\label{sec:conclusion}

In this work, we propose a feedback-driven iterative framework for synthesizing high-quality test cases. By establishing a closed-loop process that leverages execution results from known correct and incorrect solutions as feedback signals, our method systematically refines test case generators to achieve both high fidelity (TPR) and strong discriminative power (TNR). Applying this framework to competitive programming, we construct CodeContests-O, which significantly outperforms existing datasets and yields substantial performance gains when used as reward signals in reinforcement learning. These results confirm that iterative refinement guided by execution feedback is essential for distinguishing genuine logical reasoning from superficial pattern matching. 

Beyond competitive programming, our framework generalizes to any domain where problems are paired with verifiable correct and incorrect solutions, such as mathematical reasoning, formal theorem proving, and constraint satisfaction tasks, offering a general recipe for constructing high-fidelity verifiable datasets across diverse domains.

\section*{Limitations}

While our framework significantly improves test case quality, the primary limitation is the computational overhead associated with the iterative process. Reaching the optimal Pareto front requires multiple rounds of model inference and code execution to refine the generator, checker, and command sets, resulting in higher latency compared to single-pass methods. Additionally, our approach is designed to augment existing programming problems by generating high-quality test cases; it does not generate new problem statements from scratch.


\bibliography{references}

\clearpage
\appendix

\section{CodeContests-O Dataset Details}
\label{apx:dataset_details}

\subsection{Problem Selection and Filtering}
\label{sub_apx:problem_filter}

To ensure a high-quality problem set, we implement a series of heuristic rules to filter the initial problem pool $\mathcal{D}$. Specifically, A problem $\mathcal{P}$ is excluded from the dataset if it satisfies any of the following criteria:

\begin{itemize} 
    \item \textbf{Incomplete Descriptions}: Problems that lack a formal text description or contain severely fragmented information, making it impossible to extract the underlying logic. 
    \item \textbf{Absence of Reference Solutions}: Problems without any verified reference solution $S^*$, which is essential for the initial validation of synthesized test cases. 
    \item \textbf{Multimodal Inputs}: Problems requiring the processing of non-textual information, such as images or diagrams. 
    \item \textbf{Non-Standard Execution Formats}: Problems that do not require a complete, standalone program, such as (i) \textit{Function-only tasks}, where the implementation is restricted to a specific function within a predefined framework, or (ii) \textit{Interactive problems}, which necessitate real-time communication with an external grader during execution. These cases are excluded to ensure a uniform and automated environment.
\end{itemize}

\subsection{Dataset Distribution and Statistics}
\label{sub_apx:dataset_statistics}

CodeContests-O builds upon the original CodeContests dataset, with its primary improvements focused on the quality of test cases and solution pools:

\begin{itemize}
    \item \textbf{Core Refinement:} Our method focuses on generating high-quality test cases $\mathcal{T}^{e}$ and filtering the solution pools ($S^+$ and $S^-$). This process ensures higher reward fidelity and discriminative power by eliminating noise and non-compilable entries.

    \item \textbf{Property Inheritance:} The remaining information for each problem is essentially inherited from CodeContests. This includes all other problem-level metadata, such as time/memory limits, difficulty ratings, and tags.
\end{itemize}

\section{Additional Details of Feedback-Driven Iterative Framework}
\label{apx:method_details}

\subsection{Context Compression Mechanism}
\label{sub_apx:context_compression}

To maintain computational efficiency and avoid exceeding the model's context window during the iterative refinement process, we implement a context compression strategy. In a standard multi-turn refinement, the dialogue history accumulates as follows: problem statement $\rightarrow$ initial candidates of generator and commands $\rightarrow$ feedback on execution results $\rightarrow$ refined generator and commands.

To optimize this, we compress the multi-turn interaction into a consolidated representation. Once the generator and command sets are refined based on execution feedback, we discard the intermediate, suboptimal iterations and the granular feedback logs. Instead, we restructure the context to pair the original problem statement directly with the latest refined generator and commands.

By treating the refined output as if it were the response to the initial problem, we effectively "collapse" the conversation history. This ensures that the model only retains the most potent reasoning traces and final high-fidelity outputs, significantly reducing the token overhead while preserving the essential logic required for further iterations.

\subsection{Synchronized Checker Generation and Refinement}
\label{sub_apx:chekcer_generation}

To ensure high-fidelity evaluation for problems with multiple valid outputs, we implement a co-generation strategy for the checker. Rather than using static string matching, the model synthesizes the checker and the generator within the same process. This ensures the checker is inherently aligned with the specific constraints of the generated test cases. Throughout the refinement loop, the checker is optimized alongside the generator and command sets, allowing its verification logic to adapt to increasingly complex edge cases.

This logic-based checker takes the test input as context to evaluate the consistency between the solution output and the reference output. By analyzing logical correctness rather than raw text, it accurately identifies valid solutions even when multiple correct paths exist. This simultaneous optimization of test cases and verification logic significantly boosts the dataset's discriminative power, ensuring both rigorous and fair evaluation across diverse problem types.

\subsection{Prompt Templates}
\label{sub_apx:prompt_templates}

Here we provide the complete prompt templates used in our feedback-driven iterative framework. Prompt~\ref{prompt:initial_test_cases_generation} presents the template for the Initial Test Case Generation phase, while Prompt~\ref{prompt:feedback_guided_refinement} details the template for Feedback-Guided Refinement. Our generators are implemented using testlib\footnote{\url{https://github.com/MikeMirzayanov/testlib}}, a standard library widely adopted by Codeforces for creating contest problems. As the checker follows the same generation and refinement process as the generator, we omit its prompts for brevity.

Since CodeContests+ has already synthesized preliminary generators for each problem, we leverage these existing generators as a starting point rather than generating from scratch to reduce computational overhead. Our prompts instruct the LLM to analyze and improve upon these generators through a search-and-replace mechanism.

\section{LLM Usage}
LLM assistance was used exclusively for linguistic refinements, including grammar corrections and wording improvements, as well as assisting with visualization. The LLM was not involved in research ideation, experimental design, data analysis, or substantive content creation. All intellectual contributions remain solely the authors' work.

\onecolumn
\addtocounter{section}{-1}
\begin{promptbox}[Initial Test Case Generation Prompt]{prompt:initial_test_cases_generation}
You are an expert in generating command-line arguments for corner case generation programs for programming problems.

Given the following problem statement and a C++ generation program, your tasks are:
1. Carefully read and understand the problem statement.
2. Carefully read and understand the provided generation program, which is designed to generate corner case inputs for this problem.
3. Identify and summarize the constraints of the input data.
4. Analyze the problem and the generation program to anticipate common mistakes or edge cases that contestants might overlook.
5. If the provided generator is incomplete or insufficient to produce high-quality adversarial cases (e.g., missing modes/flags/branches or has buggy logic), propose minimal, concrete generator code improvements using search-replace blocks. Each block must strictly follow the pattern:
    <<<<<<< SEARCH
    <original code fragment to search for>
    =======
    <replacement fragment (the improved code)>
    >>>>>>> REPLACE
    Notes:
    - Provide only the smallest necessary surrounding context to uniquely match; avoid large blocks.
    - Prefer multiple small, focused replacements over a single massive one.
    - Do not add explanations around the blocks; return only the blocks themselves as strings.
    - Pay close attention to code indentation, spaces, and line breaks; do not omit or alter them in the search/replace fragments.
    - For each SEARCH block, you must strictly copy the exact content from the provided generator. Do NOT add or modify any characters, such as adding "-" or "+" at the beginning of lines. The SEARCH block must be an exact substring of the generator.
    - For each REPLACE block, strictly follow the code format and ensure that after replacing the SEARCH content with the REPLACE content, the generator can be compiled and run directly.
    - In the REPLACE blocks you add, if you need to introduce new functions or variables, ensure that these functions or variables are already defined or imported in the generator. Do not introduce non-existent functions or variables, and carefully check whether the parameters of the called functions are correct. For example, the common `rnd.shuffle()` function may cause an error: 'class random_t' has no member named 'shuffle'; `rnd.next` requires two arguments of the same type; the `ensure` function only accepts one argument, etc. Below is the header comment from the testlib.h used by the code:
/*
 * It is strictly recommended to include "testlib.h" before any other include
 * in your code. In this case testlib overrides compiler specific "random()".
 *
 * If you can't compile your code and the compiler outputs something about
 * ambiguous calls of "random_shuffle", "rand" or "srand", it means that
 * you shouldn't use them. Use "shuffle", and "rnd.next()" instead because
 * these calls produce stable results for any C++ compiler. Read
 * sample generator sources for clarification.
 *
 * Please read the documentation for class "random_t" and use the "rnd" instance in
 * generators. These sample calls might be useful for you:
 *              rnd.next(); rnd.next(100); rnd.next(1, 2);
 *              rnd.next(3.14); rnd.next("[a-z]{{1,100}}").
 *
 * Also read about wnext() to generate off-center random distributions.
 *
 * See https://github.com/MikeMirzayanov/testlib/ to get the latest version or bug tracker.
 */

    - In the REPLACE blocks you add, if you need to reference variables from other parts of the code, carefully check their scope to ensure that the referenced variables are visible in the generator.
6. Based on your analysis and the improved generator, design and output a diverse set of command-line commands ("command_list") that, when executed, will use the generation program to generate corner case inputs that cover as many special and adversarial cases as possible. Note that the format and arguments of the command line must comply with the requirements of the generation program. For example, ensure that --seed may be an invalid argument, and when --n usually expects a numeric value, do not pass a string.

Problem Statement:
{problem_statement}

Generation Program (C++):
{generator}

**Strictly follow these output requirements:**
- Your response must be in JSON format matching this structure:
    {{
        "input_constraints_summary": "string describing input constraints from the problem statement",
        "search_replace_generator_blocks": [
            "<<<<<<< SEARCH\n<original>\n=======\n<replacement>\n>>>>>>> REPLACE",
            ...
        ],
        "command_list": ["./gen --arg1 value1 ...", "./gen --arg2 value2 ...", ...]
    }}
- The "input_constraints_summary" field should contain a clear and concise summary of all input constraints, including both explicit constraints mentioned in the problem statement (such as input size limits, value ranges, format requirements, etc.) and any implicit constraints that can be inferred from the problem description (such as properties, invariants, or hidden requirements implied by the problem context).
- `search_replace_generator_blocks` is optional-include it only when the generator needs improvements. Each item must strictly follow the search-replace block format shown above. If no changes are needed, return an empty list ([]). If changes are proposed, ensure that `command_list` is generated against the updated generator (i.e., after applying the edits).
- The "command_list" field must contain a list of shell commands, each starting with './gen' and followed by the appropriate arguments for the generation program. Each command should be designed to generate one corner case input. All corner case inputs generated by these commands should be as diverse and adversarial as possible, covering a wide range of edge cases and adversarial scenarios.
- Do not generate the corner case inputs directly; only generate the command lines to run the generation program.
- The commands should be ready to execute in a Linux shell and should use proper argument formatting as required by the generation program.
\end{promptbox}

\begin{promptbox}[Feedback-Guided Refinement Prompt]{prompt:feedback_guided_refinement}
Now you need to refine the previously generated command list for the corner case generation program based on evaluation feedback.

You previously generated a set of commands for the given programming problem. The process is as follows:
1. The generated `search_replace_generator_blocks` have already been applied to the generator. Any blocks whose SEARCH fragments did not match exactly were skipped.
2. Each command is executed to generate one or more corner case inputs.
3. For each generated corner case input, the canonical solution is executed to obtain the corresponding output, thus forming a complete corner case (input + output).
4. These corner cases are then used to evaluate both correct solutions and incorrect solutions.

Current improved Generation Program (C++) (Note: The edits from the previously returned `search_replace_generator_blocks` have already been applied to the generator below. Any blocks that are not reflected were skipped because their SEARCH fragments did not match exactly. If the previous `search_replace_generator_blocks` was empty or none of the blocks were applied, then the generator shown here is the same as the originally provided generator and will appear as an empty string):
{improved_generator}

Current command list: {current_command_list}

For each command, here are the corresponding generated corner case input(s) (for some commands that generate very long inputs, the input for that command is replaced by `[input]`):
{command_to_input_map}

If any command failed to execute or produced errors when generating input, here are the error messages (if any):
{command_run_errors} (ideally, this should be empty)

The evaluation results are as follows (ideally, all three should be empty):
- Outputs from correct solutions: These are cases where the generated corner cases incorrectly cause correct solutions to fail (i.e., the correct solution is judged as wrong on these cases). {correct_results}
- Outputs from incorrect solutions: These are cases where the generated corner cases fail to expose bugs in incorrect solutions (i.e., the incorrect solution is judged as correct on these cases). {incorrect_results}
- Outputs from the canonical solution (only includes results for cases that failed when run with the canonical solution): These are cases where the canonical solution itself fails or produces errors on the generated corner cases. {outputs}

Please note:
- For some commands that generate very long inputs, the `stdin` field in `correct_results`/`incorrect_results`/`outputs` may be replaced by the corresponding command string, and the field will have a trailing ` [command]` tag to indicate this substitution. When you see such a `stdin` value, you should use the provided mapping between commands and generated inputs to implicitly convert the command back to its actual `stdin` content for any reasoning, comparison, or decision-making tasks.
- For some cases where the output (`stdout`/`expected_output`) is very long, the `stdout`/`expected_output` field may be replaced by `[output]`/`[expected output]`. When you see `[output]`/`[expected output]`, in this case, if the solution's `passed` field is False, you should rely only on the given solution content for reasoning.

Here is a clear and concise summary of the input constraints mentioned in the problem statement (e.g., input size limits, value ranges, format requirements, etc.): {input_constraints_summary}

Your tasks are:
1. Based on the above canonical solution results, identify any commands that generate invalid or unhelpful corner cases (i.e., those that fail when run with the canonical solution) and mark them for replacement.
2. Based on the correct solutions results, identify commands that generate corner cases which incorrectly classify correct solutions as wrong, and mark them for replacement.
3. Analyze the above results to determine which commands fail to effectively distinguish between correct and incorrect solutions.
4. If the provided generator is incomplete/insufficient to produce high-quality adversarial cases (e.g., missing modes/flags/branches or has buggy logic), propose minimal, concrete generator code improvements using search-replace blocks. 
5. Generate new additional commands that can better expose bugs in incorrect solutions and improve differentiation between correct and incorrect solutions.

**Strictly follow these output requirements:**
- Your response must be in JSON format matching this structure:
    {{
        "search_replace_generator_blocks": [
            "<<<<<<< SEARCH\n<original>\n=======\n<replacement>\n>>>>>>> REPLACE",
            ...
        ],
        "replace_command_list": ["old_command_1", "old_command_2", ...],
        "add_command_list": ["new_command_1", "new_command_2", ...]
    }}
- `search_replace_generator_blocks` is optional-include it only when the generator needs improvements. Each item must strictly follow the search-replace block format shown above. If no changes are needed, return an empty list ([]). If changes are proposed, ensure that both `replace_command_list` and `add_command_list` are generated against the updated generator (i.e., after applying the edits).
- `replace_command_list` contains commands from the original list that should be removed/replaced due to generating invalid or unhelpful corner cases, or incorrectly classifying correct solutions as wrong.
- `add_command_list` contains new commands to be added to better distinguish correct and incorrect solutions, including improved versions of replaced commands and completely new adversarial commands.
- Each command should be a shell command starting with './gen' and followed by the appropriate arguments for the generation program.
- Do not generate the corner case inputs directly; only generate the command lines to run the generation program.
- The commands should be ready to execute in a Linux shell and should use proper argument formatting as required by the generation program.

Please focus on maximizing the adversarial value of the generated corner cases based on the feedback above.
\end{promptbox}

\end{document}